\documentclass[fleqn,twoside]{article}
\usepackage{graphicx}
\usepackage{amsmath,amssymb}
\usepackage{espcrc2}
\usepackage[pdftex]{xcolor}
\newlength{\graphicwidth}
\newlength{\graphiclength}

\title{Model-independent analysis of $\tau \rightarrow \ell \ell \ell'$ decays}
\author{Sascha~Turczyk\address[Usiegen]{Theoretische Physik 1, Fachbereich Physik, Universit\"at Siegen, D-57068 Siegen, Germany}}

\begin{document}
\begin{abstract}
Many models for physics beyond the Standard Model predict lepton-flavour violating 
decays of charged leptons at a level which may become observable very soon. We investigate 
the decays of a $\tau$-lepton into three charged leptons ($\tau \rightarrow \ell \ell \ell' $, $\ell^{(}{'}^{)} = e,\mu$) in a generic way. Using effective-field-theory methods, the relevant operators are classified according to their chirality structure.
For each case, we work out the Dalitz plots for the decay distributions, including interference terms which arise from
four-lepton operators and radiatively induced processes. We discuss phenomenological implications, in particular the potential to distinguish different new physics models.
\end{abstract}
%
\maketitle
\section{Introduction}
Over the last few years lepton flavour violation (LFV) has become a hot topic. On the one hand, neutrino oscillations have been discovered recently, which is already a clear signal for physics beyond the Standard Model (SM). Incorporating these effects into a minimal extension of the SM, LFV for the charged leptons is predicted, but however at a completely unobservable level. On the other hand, many models forcast measurable new physics (NP) effects at the TeV scale, which can be investigated soon with the upcoming experiments. These extensions of the SM predict LFV at much higher rates, which, in some cases, may already be in conflict with existing experimental bounds \cite{PDG} (see also \cite{Banerjee:2007rj} for a recent summary of $B$\/-factory results).
With the advent of new experimental facilities \cite{Yamada:2005tg} (see also \cite{SuperB}) the current bounds will be pushed further, if not a discovery will be made. 

To identify effects beyond the SM one needs, among others, reliable bounds on LFV decay modes. This requires knowledge on the phase space dependence. Furthermore if a discovery is made, one can directly compare different NP models. But due to the huge number of different models, this is quite cumbersome. On the other hand a model-independent approach allows us to classify the effects in a general way. This results can then in principle be used to compare or restrict different models.

All this models which predict LFV $\tau$ decays of the form $\tau \to \ell \ell' \ell '' $ with $\ell, \ell', \ell '' = e, \mu$
\cite{Ilakovac:1994kj,Barbieri:1995tw,Hisano:1995cp,Ellis:2002fe,Brignole:2004ah,Masiero:2004js,Arganda:2005ji,Paradisi:2005fk,Chen:2006hp,Goyal:2006vq,Choudhury:2006sq,BurasBlanke} will eventually match onto a set of local four-fermion operators or radiative operators, the latter mediating $\tau \to \ell \gamma^*$ with subsequent decay of the (virtual) photon into a charged lepton pair.

We persue a bottom-up approach, where we regard the SM as an effective-field-theory (EFT). The NP effects are expanded in terms of $1/\Lambda$, where $\Lambda$ is the scale corresponding to the NP scenario. This allows us to consider all possible four-fermion and radiative operators with arbitrary coupling constants, which can be determined by studying the decay distributions of the three leptons in the final state. NP models are then distinguished by using the fact that in different models different operators dominate. Even if no signal events are found, such a study of the decay distributions is necessary to determine the efficiency of an experiment and hence to extract reliable limits. 

At the LHC experiments it will be possible to detect LFV decays of a $\tau$ lepton, especially into channels with three leptons. The signal $\tau \to 3 \mu $ will be one of the cleanest signatures \cite{LHCexp}; we will therefore concentrate on this in the following.

First we classify the general set of effective operators and formulate the corresponding interaction theory. After pointing out the kinematical variables, we show the results for the decay distributions. In the end we shortly comment on the unknown couplings in minimal flavour violating scenarios and present a conclusion.
\section{Theoretical Basics}
\subsection{Operator Analysis}
We construct the EFT by assuming that some contribution from unknown physics beyond the SM at a high scale $\Lambda$ induces LFV processes.\footnote{Notice that, in general, the scale associated to lepton-flavour violation is independent
of the scale related to lepton-number violation, $\Lambda_\text{LN}$.} At the electro-weak scale these LFV interactions manifest themselves in higher dimensional operators, which have to be compatible with the $SU(2)_L \times U(1)_Y$ gauge symmetry of the SM. 
The leading operators for processes involving charged leptons will be of dimension 6.\footnote{Dimension 8 operators, which involve scalar and tensor currents, are further suppressed by $v^2/\Lambda^2$. The suppression in 2HDM could be reduced due to large $\tan \beta$ effects.}
We will construct these operators in the following.

To respect the SM gauge symmetry we group the left-handed leptons in a $SU(2)_L$ doublet, while the right-handed charged leptons (which are singlets under $SU(2)_L$) are put into an incomplete doublet, as a reminiscent of a right-handed $SU(2)_R$ related to custodial symmetry.
Writing also the Higgs boson in matrix form, we have
\begin{align}
	L &=  \left(\begin{array}{c} \nu_L \\ \ell_L \end{array} \right) \, ,  \quad R = \left(\begin{array}{c} 0 \\ \ell_R \end{array} \right)  \, , \nonumber\\
	H &= \frac{1}{\sqrt{2}}   \left( \begin{array}{cc}   v+h_0 + i \chi_0   &\   \sqrt{2} \phi_+ \\  - \sqrt{2} \phi_-   &\  v+h_0 - i \chi_0 \end{array} \right)\,. \label{fields}
\end{align} 
For simplicity we have suppressed the family indices, which will be specified once we consider a particular decay mode.
In terms of the fields defined in (\ref{fields}),
the list of operators relevant for the lepton flavour violating $\tau \to \ell\ell'\ell''$ decays
reads
\begin{subequations}
\begin{align}
	&\textbf{dimension 6 leptonic:}\nonumber \\
	&O_1 = (\bar{L} \gamma_\mu L)  (\bar{L} \gamma^\mu L)  \label{O1}\\
	&O_2 = (\bar{L} \tau^a  \gamma_\mu L)  (\bar{L} \tau^a \gamma^\mu L)  \\
	&O_3 =  (\bar{R} \gamma_\mu R)  (\bar{R} \gamma^\mu R) \\
	&O_4 =  (\bar{R} \gamma_\mu R)  (\bar{L} \gamma^\mu L) \label{O4}\\[0.25\baselineskip]
	&\textbf{dimension 6 radiative:} \nonumber \\
	&R_1 =  g' (\bar{L} H  \sigma_{\mu \nu}  R) B^{\mu \nu}  \label{R1}\\
	&R_2 =  g (\bar{L}  \tau^a H \sigma_{\mu \nu}   R) W^{\mu \nu, a} \label{R2}
\end{align}
\end{subequations}
where $B_{\mu \nu}$ is the $U(1)_Y$ gauge field, $W_{\mu \nu}^a$ are the
$SU(2)_L$ gauge fields and $g$ and $g'$ are the corresponding gauge couplings. In this analysis we concentrate only on operators which contribute at tree-level; operators which are bi-linear in the lepton fields
and contribute at the loop-level, can be found e.g.\ in \cite{Hisano:1995cp,Brignole:2004ah,Cirigliano:2005ck}. 
The most general effective Hamiltonian at the electro-weak scale is then obtained by summing over these operators, multiplied by arbitrary coefficients for every flavour combination. In a particular new physics scenario, these coefficients should be obtained by matching at the new physics scale $\Lambda$ and evolving down to the scale $M_W$ within the
SM which is considered as an effective theory (ET).

To describe LFV decays of a $\tau$ lepton into three charged
leptons, we have to construct the effective interaction at the scale of the
$\tau$ lepton. To this end we integrate out the weak gauge bosons and the Higgs. In the following we will focus on $\tau^-$ decays; the decay distributions for $\tau^+$ decays are identical. On the level of four-fermion operators with dimension 6, we obtain the same structures as in (\ref{O1}-\ref{O4}). Projecting on charged
leptons only, we see that $O_2$ becomes equivalent to $O_1$, and both
match onto a purely left-handed operator
\begin{equation} \label{LLLL}
	H_\text{eff}^{(LL)(LL)} = g_V^{(LL)(LL)} \, \frac{(\bar{\ell}_L \gamma_\mu \tau_L )(\bar{\ell}'_L \gamma^\mu \ell''_L)}{\Lambda^2} \,,
\end{equation}
where here and in what follows the superscript of the coupling denotes the
combinations of chiralities involved and the subscript denotes the relevant
Dirac structure.
Likewise, the operator $O_3$ corresponds to a purely right-handed interaction
\begin{equation} \label{RRRR}
	H_\text{eff}^{(RR)(RR)} = g_V^{(RR)(RR)} \, \frac{(\bar{\ell}_R \gamma_\mu \tau_R )(\bar{\ell}'_R \gamma^\mu \ell''_R)}{\Lambda^2} \,,
\end{equation}
while we get a mixed term from the operator $O_4$
\begin{align} 
	H_\text{eff}^{(LL)(RR)} = &g_V^{(LL)(RR)} \, \frac{(\bar{\ell}_L \gamma_\mu \tau_L )(\bar{\ell}'_R \gamma^\mu \ell''_R)}{\Lambda^2} \nonumber\\
	+ &g_V^{(RR)(LL)} \, \frac{(\bar{\ell}_R \gamma_\mu \tau_R )(\bar{\ell}'_L \gamma^\mu \ell''_L)}{\Lambda^2} \,.\label{LLRR}
\end{align}

In case of the dimension 6 radiative operators we are only interested in the neutral current component coupling to a charged lepton pair, which mediates the LFV decay into three charged leptons. Switching to the physical fields accounting for the neutral current component, the photon and $Z_0$ fields, we have to integrate out both the Higgs and the $Z_0$. In doing so we obtain a radiative operator with a photon, as well as a four-fermion contribution from $Z_0$ exchange. We find
that this operator related to $Z_0$ exchange is suppressed relative to the photon exchange and four-fermion operators by the small Yukawa coupling of the $\tau$ lepton.  Thus, only the photonic contribution has to be taken
into account. For this we obtain at the scale $m_\tau$
\begin{equation} \label{rad}
	H_\text{eff}^\text{rad} = \frac{e}{4 \pi} \, \frac{v}{\Lambda^2} \, \sum_{h,s}  g_\text{rad}^{(s,h)} \left(\bar{\ell}_h (-i \sigma_{\mu\nu})  \tau_s \right) F^{\mu \nu}\,,\!\! 
\end{equation}
where $g_\text{rad}^{(L,R)}$ and $g_\text{rad}^{(R,L)}$ 
denote the two possible chirality combinations.\footnote{
Here, for simplicity, we neglected possible form factor effects  for
decays into virtual photons from long-distance lepton or quark loops. In the
most general case, the $\tau \to \ell \gamma^*$ vertex could be parametrized as
\begin{align*}
	\frac{e}{4 \pi} \, \frac{v}{\Lambda^2} \, \sum_{h,s} \, \bar{\ell}_h \big\{ 
		&g_\text{rad}^{(s,h)}(q^2) \, (-i \sigma_{\mu\nu}) \, q^\mu  \\[-0.5\baselineskip]
		+& m_\tau \, f_\text{rad}^{(s,h)}(q^2) \, \left(\gamma_\nu - \frac{q_\nu}{q^2} \, \slash q \right)
	\big\}  \tau_s \,,
\end{align*}
where $g_\text{rad}^{(s,h)}(0) \equiv g_\text{rad}^{(s,h)}$ 
and $f_\text{rad}^{(s,h)}(0)=0$, see e.g. \cite{Raidal:1997hq}.}
The matrix element for $\tau \to \ell  \bar{\ell}' \ell'$ becomes
\begin{align}
	&\langle \ell \bar{\ell}' \ell' | H_\text{eff}^\text{rad} | \tau \rangle = \alpha_{em}   \,  \frac{v}{\Lambda^2} \, \frac{q^\nu}{q^2}\, \nonumber\\
	\times &\!\sum_{h,s}  g_\text{rad}^{(s,h)}  \, \langle \ell \bar{\ell}' \ell' | \left(\bar{\ell}_h (-i \sigma_{\mu\nu})  \tau_s\right) \left(\bar{\ell}' \gamma^\mu \ell'\right)| \tau \rangle \,,\label{radhad}
\end{align}
where $q$ is the momentum transfer through the photon. This momentum transfer
is proportional to the lepton masses, and thus this contribution scales as
$1/(y \Lambda^2)$ where $y$ is a Yukawa coupling of the leptons, which would
lead to an enhancement unless an additional Yukawa coupling appears in the
numerator as, for instance, in minimal flavour violation (MFV).

Four different Dirac structures can occure if only the four-lepton operators contribute. Two additional Dirac structures are possible for decays of the form $\tau^- \to {\ell'}^- \ell^- \ell^+$, where also the radiative operator has to be taken into account.

The corresponding coupling constants
\[
	g_V^{(LL)(LL)}, \quad g_V^{(RR)(RR)}, \quad  g_V^{(LL)(RR)}, \quad g_V^{(RR)(LL)}
\]
of the four-lepton operators and additionally
\[
	g_\text{rad}^{(LR)}, \quad g_\text{rad}^{(RL)} 
\]
from the radiative operators are matrices in lepton flavour space.
\subsection{Decay Modes}
There are in total six different decay modes of the $\tau^-$ to consider
\begin{subequations}
\begin{align}
	\tau^- &\to  e^- e^- e^+\label{decay1}\\
	\tau^- &\to \mu^-  \mu^- \mu^+ \label{decay2}\\
	\tau^- &\to  e^- e^- \mu^+\label{decay3} \\
	\tau^- &\to \mu^-  \mu^-  e^+\label{decay4}\\
	\tau^- &\to  \mu^- e^-  e^+\label{decay5}\\
	\tau^- &\to e^-  \mu^- \mu^+\,.\label{decay6}
\end{align}
\end{subequations}
Notice that (\ref{decay1} - \ref{decay4}) contain two identical
particles ($e^-e^-$ or $\mu^-\mu^-$) in the final state for which we have to take into account the Pauli principle (PP), whereas
(\ref{decay5} + \ref{decay6}) do not.
Moreover, only (\ref{decay1}, \ref{decay2}, \ref{decay5}, \ref{decay6})
receive contributions from the radiative operators \eqref{rad} via
\begin{equation*}
	\tau^- \to \ell^- \gamma^* \to \ell^- (\ell'{}^+ \ell'{}^-) \,.
\end{equation*}
Therefore we calculate the Dalitz distributions \(\text{d}^2\Gamma/(\text{d}m_{\!+-}^2 \!\text{d}m_{\!--}^2)\). We neglect interference terms with different helicities, because of the suppression due to small lepton masses, and take into account only the leading interference term.
\section{Results}
\setlength{\graphicwidth}{0.65\columnwidth}
\setlength{\graphiclength}{-2.05\baselineskip}
\begin{figure}[htb]
	\includegraphics[width=\graphicwidth]{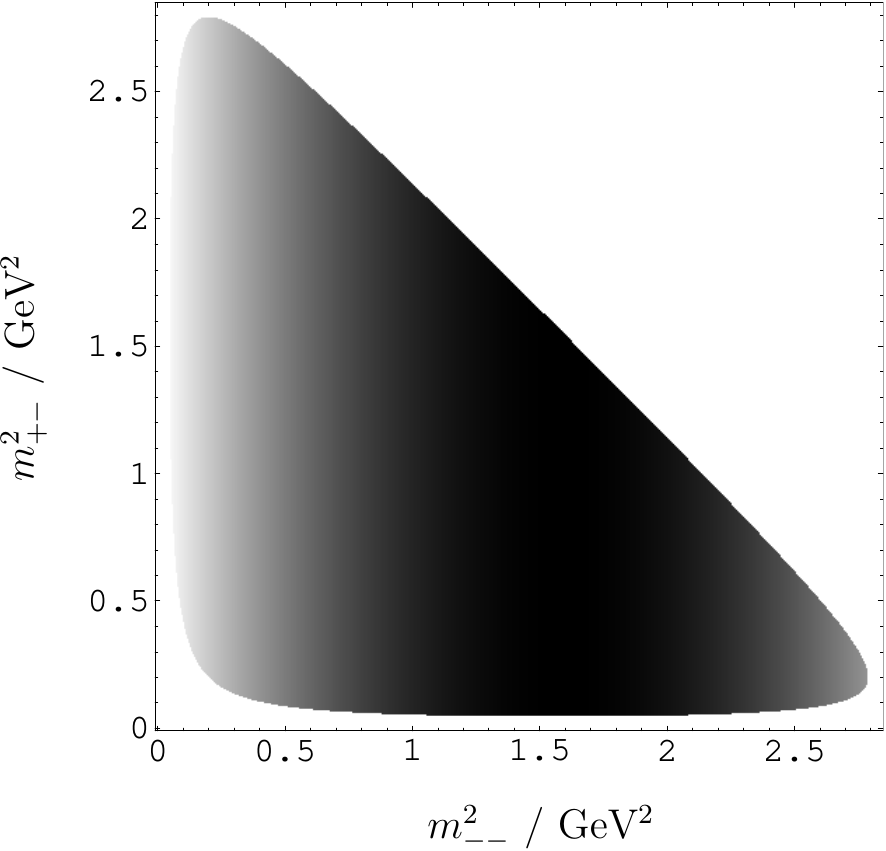}\vspace{\graphiclength}\caption{Dalitz Distribution for \(\frac{\text{d}^2\Gamma_V^{(LL)(LL)}}{\text{d}m_{23}^2 \, \text{d}m_{12}^2} \).}\vspace{\graphiclength}\label{vmmm_llll}
\end{figure}
\begin{figure}[htb]
	\includegraphics[width=\graphicwidth]{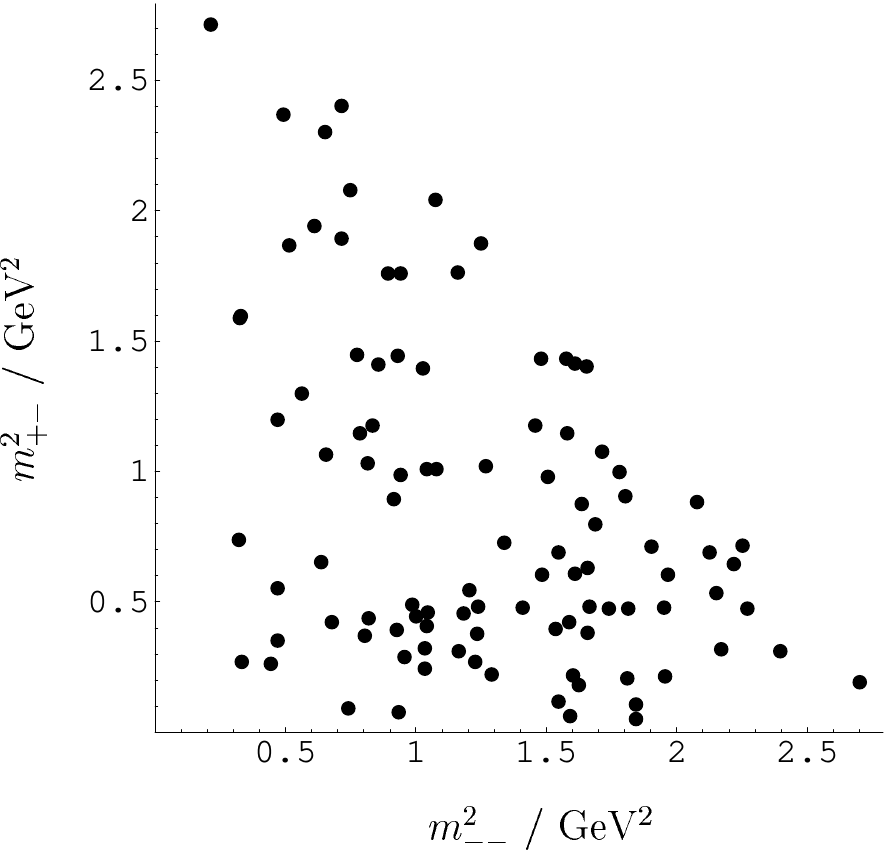}\vspace{\graphiclength}\caption{Simulated 100 events for \(\frac{\text{d}^2\Gamma_V^{(LL)(LL)}}{\text{d}m_{23}^2 \, \text{d}m_{12}^2} \).}\vspace{2.2\graphiclength}\label{vmmm_llll_100}
\end{figure}
\begin{figure}[htb]
	\includegraphics[width=\graphicwidth]{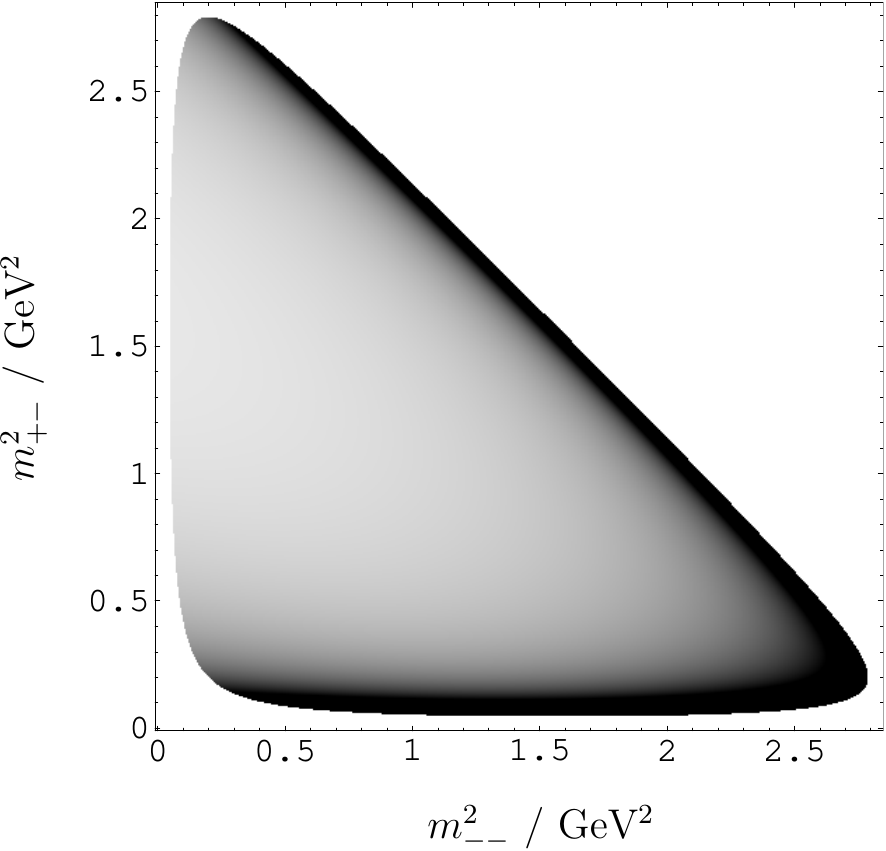}\vspace{\graphiclength}\caption{Dalitz Distribution for \(\frac{\text{d}^2\Gamma_\text{rad}^{(LR)}}{\text{d}m_{23}^2 \, \text{d}m_{12}^2} \).}\vspace{0.1\graphiclength}\label{vmmm_rad}
\end{figure}
\begin{figure}[htb]
	\includegraphics[width=\graphicwidth]{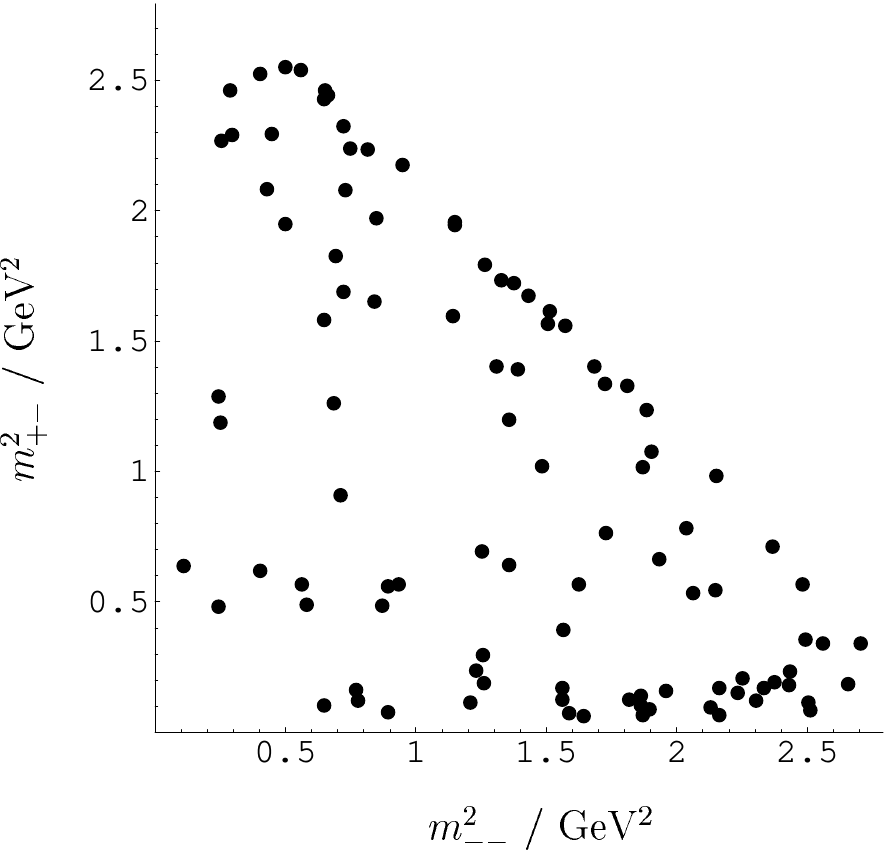}\vspace{\graphiclength}\caption{Simulated 100 events for \(\frac{\text{d}^2\Gamma_\text{rad}^{(LR)}}{\text{d}m_{23}^2 \, \text{d}m_{12}^2} \).}\vspace{\graphiclength}\label{vmmm_rad_100}
\end{figure}
The Dalitz distributions are visualised in form of density plots, made in such a way that the highest value is displayed as black. Here we concentrate on the difference between four-lepton and radiative operators and show only  the results to $\tau^- \rightarrow \mu^- \mu^- \mu^+$. Details to all different possibilities can be found in \cite{Dassinger:2007ru}. The kinematical variables are then given by
\begin{align*}
	m_{--}^2 &\equiv m_{12}^2 = (p_{\mu^-}' + p_{\mu^-})^2 \\
	m_{+-}^2 &\equiv m_{23}^2 = (p_{\mu^-} + p_{\mu^+})^2 \,.
\end{align*}
The third combination is given by
\begin{equation*}
	m_{13}^2 = m_\tau^2+ 3 m_\mu^2 - m_{--}^2-m_{+-}^2\,.
\end{equation*}
The calculation of the decay distribution for the vector current, where all particles are left-handed, results in
\begin{align}
	&\frac{\text{d}^2\Gamma_V^{(L L)(L L)}}{\text{d}m_{23}^2 \, \text{d}m_{12}^2} = \frac{| g_V^{(L_\mu L^\tau)(L_\mu L^\mu )}|^2}{\Lambda^4}\nonumber\\
	&\times \frac{(m_\tau^2-m_\mu^2)^2 - (2 m_{12}^2 -m_\tau^2-3m_\mu^2)^2}{256 \, \pi ^3\, m_{\tau }^3}\,. \label{vector}
\end{align}
The distribution \eqref{vector} for the vector current with all particles left-handed is displayed in fig.~\ref{vmmm_llll} in form of a density plot. Furthermore a simulation of 100 events based on \eqref{vector} can be found in fig.~\ref{vmmm_llll_100}. One finds a rather flat distribution over the allowed phase space region. For the radiative transition we find the distribution 
\begin{align}
	&\frac{\text{d}^2\Gamma_\text{rad}^{(L R)}}{\text{d}m_{23}^2 \text{d}m_{12}^2} =
	\alpha_{\text{em}}^2 \, \frac{|g_\text{rad}^{(L_\mu R^\tau)}|^2 \, v^2}{\Lambda^4} \bigg[\frac{2m_{12}^2- 3 m_\mu^2}{128 \, \pi^3  \, m_{\tau }^3}\nonumber \\
	&+ \frac{m_\mu^2 \, (m_\tau^2 - m_\mu^2)^2}{128 \, \pi^3 \, m_{\tau }^3} \left( \frac{1}{m_{13}^4} + \frac{1}{m_{23}^4} \right) \nonumber \\
	&+\frac{m_\mu^2 (m_\tau^4 - 3 m_\tau^2 m_\mu^2+2 m_\mu^4)}{128 \, \pi^3 \,m_{13}^2 \, m_{23}^2 \, m_{\tau }^3} + (m_{13}^2+m_{23}^2)\nonumber \\ 
	&\times \frac{m_{12}^4+m_{13}^4+m_{23}^4-6 m_\mu^2(m_\mu^2+m_\tau^2)} {256 \, \pi^3 \,m_{13}^2 \, m_{23}^2 \, m_{\tau }^3} \bigg]\,. \label{rad_distribution}
\end{align}
The corresponding density plot can be found in fig.~\ref{vmmm_rad}. Again we have made a simulation, see fig.~\ref{vmmm_rad_100}. Due to the photon pole we find an enhancement in the corners of the phase-space for the two possible combinations (because of PP) of the $\mu^-\mu^+$ pair.

\begin{figure}[hpbt]
	\includegraphics[width=\graphicwidth]{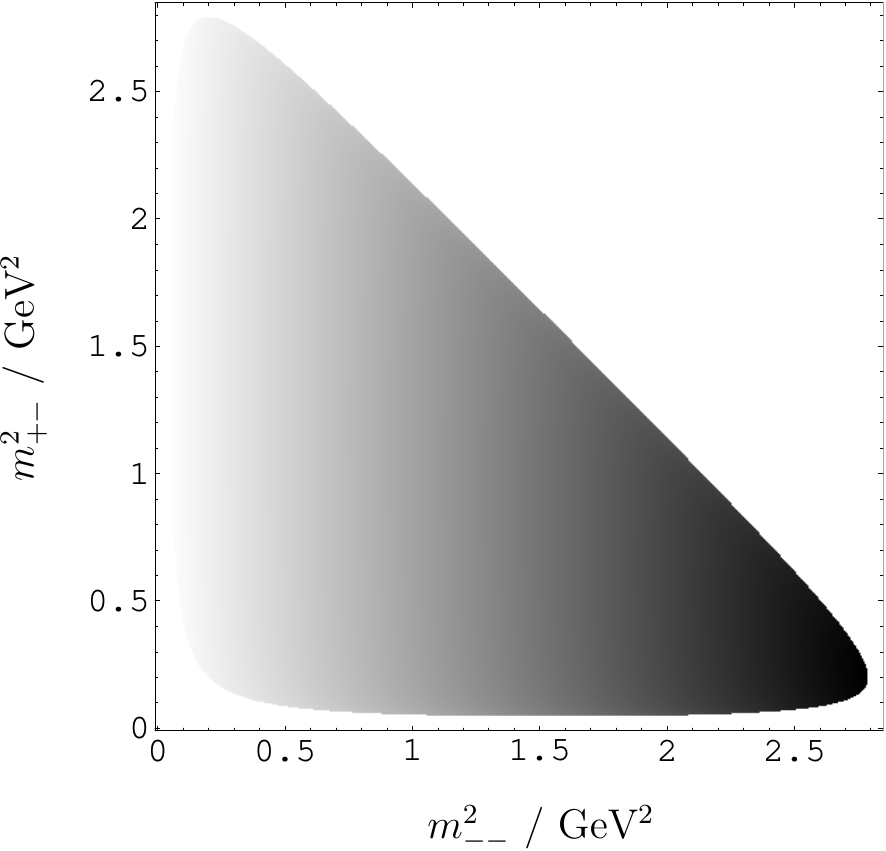}\vspace{\graphiclength}\caption{Dalitz Distribution for \(\frac{\text{d}^2\Gamma_\text{mix}^{(LL)(LL)}}{\text{d}m_{23}^2 \, \text{d}m_{12}^2} \).}\vspace{\graphiclength}\label{mixmmm_llll}
\end{figure}

\begin{figure}[ht]
	\includegraphics[width=\graphicwidth]{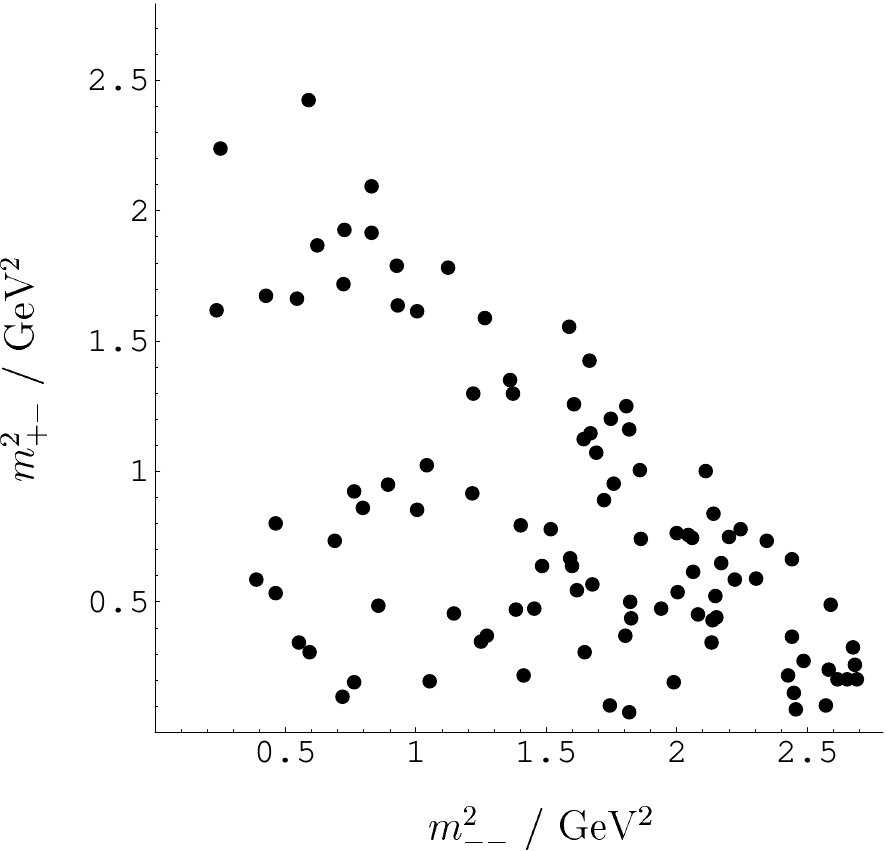}\vspace{\graphiclength}\caption{Simulated 100 events for \(\frac{\text{d}^2\Gamma_V^{(LL)(LL)}}{\text{d}m_{23}^2 \, \text{d}m_{12}^2} \).}\vspace{\graphiclength}\label{mixmmm_llll_100}
\end{figure}

Finally we have considered the leading interference term, between the radiative transition, coupling to a left-handed pair of muons and the four-lepton operator with all particles left-handed. The suppression is given by $m_\tau/v$, at least in MFV scenarios. The distribution is given in \eqref{mix} and plotted in fig.~\ref{mixmmm_llll}. The simulation can be found in fig.~\ref{mixmmm_llll_100}
\begin{align}
	&\frac{\text{d}^2\Gamma_\text{mix}^{(LL)(LL)}}{\text{d}m_{23}^2 \text{d}m_{12}^2} = \alpha_\text{em} \frac{ 2  \, v \, \text{Re}[ g_\text{V}^{(L_\mu L^\tau)(L_\mu L^\mu)} \, g_\text{rad}^{*(L_\mu R^\tau)} ]\! }{\Lambda^4} \nonumber\\
	&\times\left[ \frac{m_{12}^2-3 m_\mu^2}{64 \, \pi^3 \, m_\tau^2} + \frac{m_\mu^2(m_\tau^2-m_\mu^2)(m_{13}^2+m_{23}^2)}{128 \, \pi^3 \, m_\tau^2 \, m_{13}^2 \, m_{23}^2} \right]\,.\label{mix}
\end{align}

\subsection{Comment on MFV scenarios}
Starting point is a scenario with the minimal field content, as present in the SM. The neccessary breaking of the lepton flavour symmetry  $SU(3)_L \times SU(3)_{E_R}$ is described by two spurion fields
\begin{align*}
	\lambda &= \frac{1}{v} \, \text{diag} (m_e, m_\mu, m_\tau) \\
	g_\nu &= \frac{\Lambda_\text{LN}}{v^2} \, U^* \, \text{diag} (m_{\nu_1}, m_{\nu_2}, m_{\nu_2}) \, U^\dagger \,,
\end{align*}
where $\lambda\sim (\bar 3,3)$ describes the SM Yukawa couplings of the charged leptons, and $g_\nu\sim (\bar 6,1)$ is related to a dim-5 lepton-number violating term 
\begin{equation} \label{Maj}
{\cal L}_\text{Maj} = \frac{1}{2 \Lambda_\text{LN}} \left( N^T g N \right)\,.
\end{equation}
The quantum numbers of
\begin{equation}
	N = \left(T_3^{(R)} + \frac{1}{2} \right) H^\dagger L
\end{equation}
are vanishing under the complete SM gauge group. Both spurion fields are only parametrised by the lepton masses and the PMNS matrix $U$. To render the LFV operators gauge invariant, we have to insert appropriate numbers of these spurions. Here we are only considering the minimal insertion. For one right-handed field we need an insertion of  $\lambda$, which produces a suppression of $m_\ell/ v$. The four-lepton operators $L^i \, L^j \, L_k^* \, L_l^*$ need at least two insertions of $g_\nu$. The possible flavour structures can be read off from the reduction of the $SU(3)_L$ tensor product for $g_\nu$ and $g_\nu^\dagger$
\begin{equation}
 	\bar 6 \times 6 = 1 + 8 + 27\,.
\end{equation}
The singlet term corresponds to the trace of \(\text{tr}[g_\nu^\dagger g_\nu]\) and does not induce flavour transitions at all.
The octett \(\Delta =\Delta^\dagger  \) describes flavour transitions from 2-lepton as well as from 4-lepton operators. The 27plet combination \( G_{ij}^{kl}\) accounts only for the 4-lepton operators.
The dominating flavour coefficents for the relevant operators mediating LFV $\tau$ decays in MFV are given by
\begin{subequations}\label{coeffs}
\begin{align}
	g_V^{(L_k L^i)(L_l L^j)} &\to 2 c_1 \Delta^{k}_i \delta_{j}^{l}  + c_2   G_{ij}^{kl}  \\
	g_V^{(L_k L^i)(R_l R^j)} &\to c_3  \Delta_i^k \delta_{j}^{l} \\
	g_\text{rad}^{(L_k R^i)} &\to c_4 \Delta^k_i \lambda_i\,.
\end{align}
\end{subequations}
Therefore, assuming a MFV scenario, we can relate the couplings to the lepton masses and neutrino parameters via the relations \eqref{coeffs}.
\section{Summary and Conclusion}
LFV $\tau$ decays provides an important test of new physics contributions against the standard model, since many new physics models allow for a huge enhancement of this decay channels. Therefore a great potential to falsify the SM in the upcoming experiments is given. We have discussed LFV $\tau \rightarrow \ell \ell\ell$ processes in a model independent way. For our analysis we regard the SM as an effective theory and classify different operators within this effective theory. The operators differ in their chirality and Dirac structure. We have calculated the Dalitz distributions for these different operators including the leading interference term. It turns out that four-lepton operators show a rather uniform behaviour, whereas the decay distributions of radiative operators show large enhancement in the corners of the phase space. Therefore this analysis allows to distinguish different models with already only a few number of events, because different models give rather different predictions for the relative size of radiative and four-fermion operators. This clearly shows that using Dalitz plots for such processes is a complementary tool to the sole analysis of the total decay rates.  Additionaly the knowledge on the phase space distributions allows to improve the bounds on this decay modes by taking into account the different structures.
\section*{Acknowledgements}
This work was done in collaboration with B.~Dassinger, T.~Feldmann and T.~Mannel. S.T. thanks the organizers of the TAU 08 conference for the warm hospitality.

\end{document}